\documentclass[12pt]{article}
\usepackage{amsmath}
\usepackage{amssymb}
\usepackage{amsthm}
\usepackage{latexsym}
\usepackage{amssymb}
\usepackage{epsf}
\usepackage{mathrsfs}
\usepackage{graphicx,color}

\newtheorem{Theorem}{Theorem}

\newtheorem{Lemma}{Lemma}

\newtheorem{Condition}{Condition}

\newtheorem{Remark}{Remark}

\newcommand{\mE}{{\mathscr E}}
\newcommand{\mF}{{\mathscr F}}

\newcommand{\mK}{{\mathscr K}}

\newcommand{\mM}{{\mathscr M}}

\newcommand{\mT}{{\mathscr T}}
\newcommand{\mU}{{\mathscr U}}

\def\Dom{{\rm Dom\,}}
\def\Ran{{\rm Ran\,}}

\def\dWT{{\dot W^T}}
\def\dCT{{\dot C^T}}
\def\dFT{{\dot \mF^T}}

\def\eps{\epsilon}

\def\tH{\widetilde{\mathscr H}}

\renewcommand{\Im}{{\rm Im\,}}

\newcommand{\supp}{{\rm supp\,}}
\renewcommand{\hat}{\widehat}
\newcommand{\til}{\widetilde}
\newcommand{\ver}{\,|\,}

\newcommand{\ST}{S_0^{+,T}}
\newcommand{\SO}{S_0^+}
\newcommand{\LUT}{L_0^*\upharpoonright{\mU^T}}

\newcommand{\STX}{\Theta^{\,\xi,T}}

\def\bul{\noindent$\bullet$\,\,\,}

\numberwithin{equation}{section}

\begin{document}
\date{}
\author{
M. I. Belishev
	\thanks{Saint-Petersburg Department of V. A. Steklov Mathematical Institute, Fontanka 27, St. Petersburg 191023, Russia; belishev@pdmi.ras.ru.}\,
and 
S. A. Simonov
	\thanks{Saint-Petersburg Department of V. A. Steklov Mathematical Institute, Fontanka 27, St. Petersburg 191023, Russia; 
	St. Petersburg State University, Universitetskaya nab. 7--9, St. Petersburg 199034, Russia;
	Alferov Academic University of the Russian Academy of Sciences, Khlopina 8A, St. Petersburg 194021, Russia;
	sergey.a.simonov@gmail.com.}}

\title{A functional model of a class of symmetric semi-bounded operators}

\maketitle
\begin{abstract}
Let $L_0$ be a closed symmetric positive definite operator with nonzero defect indices $n_\pm(L_0)$ in a separable Hilbert space ${\mathscr H}$. It determines a family of dynamical systems $\alpha^T$, $T>0$, of the form
\begin{align*}
& u''(t)+L_0^*u(t) = 0  && {\rm in}\,\,\,{{\mathscr H}}, \,\,\,0<t<T,\\
& u(0)=u'(0)=0 && {\rm in}\,\,\,{{\mathscr H}},\\
& \Gamma_1 u(t) = f(t), &&0\leqslant t \leqslant T,
\end{align*}
where $\{{\mathscr H};\Gamma_1,\Gamma_2\}$\,\,\, ($\Gamma_{1,2}:{\mathscr H}\to{\rm Ker\,} L_0^*$) is the canonical (Vi\-shik) boundary triple for $L_0$, $f$ is a boundary control (${\rm Ker\,} L_0^*$-valued function of $t$) and $u=u^f(t)$ is the solution (trajectory).

Let $L_0$ be completely non-self-adjoint and $n_\pm(L_0)=1$, so that $f(t)=\phi(t)e$ with a scalar function $\phi\in {L_2(0,T)}$ and $e\in{\rm Ker\,} L_0^*$. Let the map $W^T: \phi\mapsto u^f(T)$ be such that $C^T=(W^T)^*W^T=\mathbb I+K^T$ with an integral operator $K^T$ in ${L_2(0,T)}$ which has a smooth kernel. Assume that $C^T$ an isomorphism in ${L_2(0,T)}$ for all $T>0$. We show that under these assumptions the operator $L_0$ is unitarily equivalent to the minimal Schr\"{o}dinger operator $S_0=-D^2+q$ in ${L_2(0,\infty)}$ with a smooth real-valued potential $q$, which is in the limit point case at infinity. It is also proved that $S_0$ provides a canonical wave model of $L_0$.
\end{abstract}



\section{Operators, spaces, systems}

\subsubsection*{About the paper}

\noindent\bul In the present paper we develop ideas and results of
the works \cite{BArXiv,BD,BSim_1}. Our paper is written in the
framework of the program outlined in \cite{BArXiv}. The goal of the
program is to elaborate a new functional model (the so-called
 {\it wave model}) of semi-bounded symmetric operators,
which have wide applications in mathematical physics. In
\cite{BSim_1}--\cite{BSim_5}, as well as in the present paper, the
ideas and constructions of the program are tested on concrete
classes of operators. So, before treating the most general
abstract case, we accumulate experience on examples. The class
of operators $L_0$, which we deal with here, is studied in
terms of a dynamical system with boundary control (DSBC)
determined by $L_0$.

Motivation for this activity comes from inverse problems. An
inspiring fact is that in many known cases to solve an inverse
problem is the same as to construct the wave model of a relevant $L_0$
\cite{BSim_1,BSim_3}.

\noindent\bul Note that the notation in the abstract ($W^T$, $C^T$) slightly differs from the notation that we use in the rest of the paper. Sometimes in the text we use the notation $D$ for the ordinary derivative, and $D_i$ denotes the partial derivative in $i$-th argument.

\subsubsection*{Operator $L_0$}

\noindent$\bullet$\,\,\,Let ${\mathscr H}$ be a (separable) Hilbert space,
$L_0$ a closed symmetric positive definite operator in ${\mathscr H}$, i.e., $\overline{\Dom
L_0}={\mathscr H}$, $L_0\subset L_0^*$, $(L_0y,y)\geqslant \gamma \|y\|^2$
on $\Dom L_0$ with
$\gamma>0$. We also assume that $L_0$ has nonzero defect indices
$n_+(L_0)=n_-(L_0)={\rm dim\,}\mK \geqslant 1$, where $\mK:={\rm Ker\,} L_0^*$.

By $L$ we denote the Friedrichs extension of $L_0$, so that
$L_0\subset L\subset L_0^*$ and $L\geqslant\gamma \mathbb I$ holds (see,
e.g., \cite{BirSol}). Note also that $L^{-1}$ is bounded and
defined on the whole space ${\mathscr H}$.
\smallskip

More assumptions on $L_0$ are imposed later on. The first one is the following.

\begin{Condition}\label{C1}
The operator $L_0$ is completely non-self-adjoint.\footnote{Recall
that a symmetric operator is said to be  completely non-self-adjoint,
if it has no reducing subspaces in which it has a self-adjoint part.}
\end{Condition}

\medskip

\bul The well-known decomposition by M. I. Vishik \cite{Vishik} is
$$
{\rm Dom\,}L_0^*={\rm Dom\,}L_0 \overset{.}+L^{-1}\mK\overset{.}+\mK
$$
(the sums are direct). By this, for $y \in {\rm Dom\,}L_0^*$ one has
$$
y=y_0+L^{-1}g+h
$$ 
with $y_0 \in {\rm Dom\,}L_0$ and $g,h \in
\mK$, where $h=\Gamma_1y$, $g=\Gamma_2y$ and $\Gamma_1=
L^{-1}L_0^*-\mathbb I$, $\Gamma_2=PL_0^*$, $P$ is the orthogonal projection in ${\mathscr H}$ onto $\mK$. The Green formula
\begin{equation}\label{Eq Green}
(L_0^*u, v)-(u,L_0^*v)=(\Gamma_1 u, \Gamma_2 v) - (\Gamma_2 u,
\Gamma_1 v),\qquad u,v\in\Dom L_0^*,
\end{equation}
holds \cite{BD}. The collection $\{{\mathscr H},\Gamma_1,\Gamma_2\}$ is
referred to as the {\em Vishik boundary triple} for $L_0$ \cite{DM}.

\subsubsection*{Dynamical system with boundary control}

\bul The boundary triple, in turn, determines a dynamical system
$\alpha$ of the form
\begin{align}
\label{alpha1} & u''(t)+L_0^*u(t) = 0  && {\rm in}\,\,\,{{\mathscr H}}, \,\,\,t>0,\\
\label{alpha2} & u(0)=u'(0)=0 && {\rm in}\,\,\,{{\mathscr H}},\\
\label{alpha3}& \Gamma_1 u(t) = f(t) && {\rm in}\,\,\,{\mK},\,\,\,\,\,t \geqslant 0,
\end{align}
where $f=f(t)$ is a $\mK$-valued function of time
({\em boundary control}), $u=u^f(t)$ is a solution ({\em trajectory}).

Recall that $L$ is the Friedrichs extension of $L_0$. Let
$L^{\frac{1}{2}}$ be the positive square root of $L$. For smooth\footnote{Everywhere in
the paper, {\it smooth} means $C^\infty$-smooth.} controls from
the class
$$
{\mM}\,:=\,\{f \in C^\infty\left([0,\infty);
{\mK}\right)\,|\,\,{\rm supp\,}f\subset(0,\infty)\},
$$
the solution does exist, is unique and can be represented in the form
\begin{equation}\label{Eq repres u^f}
u^f(t)\,:=\,-f(t)+\int_0^t
L^{-\frac{1}{2}}\,\sin\left[(t-s)L^{\frac{1}{2}}\right]\,f''(s)\,ds\,,
\qquad t \geqslant 0,
\end{equation}
(see \cite{BD}, where a generalized solution for a wider class of
controls is introduced). The element $u^f(t)\in{\mathscr H}$ is a state of
the system at the moment $t$. Referring to applications, we also
call $u^f$ a {\em wave}.

We always assume that all functions of time are extended to $t<0$ by zero. A
very general property of the waves is the relation
\begin{equation}\label{Eq delay rel}
u^{\mT_s f}(t)\,=\,(\mT_s u^f)(t)=u^f(t-s),\qquad s,t>0,
\end{equation}
where $\mT_s$ is the delay operator which shifts the time argument
by $t\mapsto t-s$. This follows from independence of time of
the operator $L_0^*$ that governs the evolution of the system
$\alpha$. Furthermore, the relations hold:
\begin{equation}\label{Eq delay rel+}
u^{f'}(t)\,=\,(u^f)'(t), \quad u^{f''}(t)=(u^f)''(t)=-L_0^*u^f(t),\qquad t>0.
\end{equation}
In the control and system theory relations (\ref{Eq delay
rel}) and (\ref{Eq delay rel+}) are referred to as the {\em steady-state
property} of the system $\alpha$.
\smallskip

\bul The set
$$
{\mU}^T:=\{u^f(T)\,|\,\,f \in \mM\}
$$
is called {\em reachable} (at the moment $T$), whereas
$$
{\mU}\,:=\,{\rm span\,}\{{\mU}^T\,|\,\,T>0\}
$$
is the {\em total} reachable set. The DSBC $\alpha$ is said to be
{\em controllable}, if
\begin{equation*}
\overline{\mU}\,=\,{\mathscr H}\,
\end{equation*}
holds. As shown in \cite{BD}, the system $\alpha$ is controllable
if and only if the operator $L_0$ is completely non-self-adjoint.
Thus Condition \ref{C1} provides controllability of $\alpha$.
\smallskip

\bul As one can see from (\ref{Eq repres u^f}), for any $T>0$ the wave
$u^f(T)$ is determined by the part $f\upharpoonright{[0,T]}$ of the control
(it does not depend on the values $f\upharpoonright{(T,\infty)}$). This enables one
to deal with reduced systems $\alpha^T$ of the form
\begin{align}
\label{alpha1T} & u''(t)+L_0^*u(t) = 0  && {\rm in}\,\,\,{{\mathscr H}}, \,\,\,0<t<T,\\
\label{alpha2T} & u(0)=u'(0)=0 && {\rm in}\,\,\,{{\mathscr H}},\\
\label{alpha3T}& \Gamma_1 u(t) = f(t) && {\rm in}\,\,\,{\mK},\,\,\,\,\,0\leqslant t \leqslant T.
\end{align}
The following are its attributes.

\noindent$\star$\,\,\,The {\em space of controls}
$\mF^T:=L_2([0,T];\mK)$ is called the {\it outer space} of $\alpha^T$. It
contains the family of ``delayed'' subspaces
$$
\mF^{\,T,\xi}:=\{f\in\mF^T\,|\,\,\supp f\subset[T-\xi,T]\},\qquad 0\leqslant
\xi\leqslant T.
$$
One has $\mF^{\,T,0}=\{0\}$ and $\mF^{\,T,T}=\mF^T$. The smooth class
$\mM^T:=\{f\upharpoonright{[0,T]}\,|\,\,f\in\mM\}$ is dense in $\mF^T$.
\smallskip

\noindent$\star$\,\,\,The space ${\mathscr H}$ is the {\em inner space}. It
contains the family of reachable sets $\mU^T$  and subspaces $\overline{\mU^T}$, $T>0$.
\smallskip

\noindent$\star$\,\,\,The correspondence ``input\,$\mapsto$\,state'' is
realized by the {\em control operator}
$$
W^T:\mF^T\to{\mathscr H},\quad \Dom W^T=\mM^T, \qquad W^Tf:=u^f(T).
$$
We obviously have $\Ran W^T=\mU^T$; by (\ref{Eq delay rel+}) 
the equalities
\begin{equation}\label{Eq Wd=dW}
W^Tf''=u^f_{tt}(T)=-L_0^*u^f(T)=-L_0^*W^Tf
\end{equation}
hold. As shown in \cite{DSBC_3}, the control operator is closable\footnote{but not necessarily bounded} for all $T>0$. If $W^T$ is bounded and boundedly invertible (on its image $\mU^T$), then its closure $\overline{W^T}$ has the same properties, i.e., is an isomorphism of the spaces $\mF^T$ and
$\overline{\mU^T}$. In what follows we preserve the notation $W^T$ for the closure.
\smallskip

\noindent$\star$\,\,\,The {\em connecting operator}
$$
C^T:\mF^T\to\mF^T,\quad C^T:=(W^T)^*W^T
$$
is well defined on $\{f\in\Dom W^T\ver W^Tf\in\Dom(W^T)^*\}$. By the von Neumann theorem \cite{BirSol,Kato} it is densely defined and its closure is a self-adjoint operator in $\mF^T$. We preserve the notation $C^T$ for its closure. We have $C^T\geqslant0$ and 
\begin{equation}\label{Eq C^T}
(C^Tf,g)_{\mF^T}=(W^Tf,W^Tg)_{\mF^T}=(u^f(T),u^g(T)),\quad
f,g\in\mM^T.
\end{equation}
Note that $C^T$ is bounded, if and only if $W^T$ is. Also, as one can see, $C^T$ is an isomorphism in $\mF^T$, if and only if $W^T$ is an isomorphism.
\smallskip

\noindent$\bullet$\,\,\,The steady-state relation (\ref{Eq
delay rel}) implies a corresponding property of the connecting
operator. Consider two systems $\alpha^T$ and $\alpha^{\xi}$ provided
$0<{\xi}<T$. The map
$$
\Theta^{\,\xi,T}: \mF^{\xi}\to\mF^T,\quad (\Theta^{\,\xi,T}f)(t):=
\left\{
\begin{array}{ll}
0,&0\leqslant t<T-{\xi},\\
f(t-(T-{\xi})),&T-{\xi}\leqslant t\leqslant T,
\end{array}
\right.
$$
is a partial isometry and $\mF^{\,T,{\xi}}=\Theta^{\,\xi,T}\mF^{\xi}$ holds;
the adjoint operator $(\Theta^{\,\xi,T})^*:\mF^T\to\mF^{\xi}$ acts as
\begin{equation}\label{Eq Theta*}
((\Theta^{\,\xi,T})^*f)(t)=f(t+(T-\xi)),\qquad 0\leqslant t\leqslant
\xi.
\end{equation}
Steady-state property (\ref{Eq delay rel}) easily implies
\begin{equation}\label{steady-state}
W^T\Theta^{\,\xi,T}=W^{\xi},
\end{equation}
which leads to the relation
\begin{equation}\label{Eq C^s via C^T}
C^{\xi}=(\Theta^{\,\xi,T})^*C^T\Theta^{\,\xi,T}, \qquad 0<\xi\leqslant T.
\end{equation}

\subsubsection*{Scalarized system}

\bul Now assume that in addition to Condition \ref{C1} the following is satisfied.

\begin{Condition}\label{C2}
	Defect indices of $L_0$ are $n_{\pm}(L_0)=1$.
\end{Condition} 

Then controls take the form
$f=\phi(t)e$, where $\phi\in{L_2(0,T)}=:\dot\mF^T$ is a scalar
function and $e\in\mK$, $\|e\|=1$, is a fixed nonzero element. We can regard $\dot\mF^T$ as an outer space of the system $\alpha^T$
and deal with the scalarized control
$\dot W^T:\dot\mF^T\to{\mathscr H}$, $\dot W^T\phi:=u^{\phi e}(T)$ and connecting $\dot
C^T:\dot\mF^T\to\dot\mF^T$, $\dot C^T:=(\dot W^T)^*\dot W^T$  operators. Again, $\dWT$ is an isomorphism, if and only if $\dCT$ is an isomorphism in $\dFT$.
\smallskip

\bul Assume that for every $T>0$ the scalarized connecting operator has the form
\begin{equation}\label{Eq C=I+K}
\dCT=\mathbb I+K^T,
\end{equation}
where $(K^T\phi)(t):=\int_0^Tk^T(t,s)\,\phi(s)\,ds$ with
a smooth Hermitian kernel $k^T$ \footnote{This is the case for one-dimensional inverse problems 
\cite{Blag,Blag2,B_1D_BCm,BMikh}.}. Then the kernel $k^T$
possesses the following specific property.

Consider two systems $\alpha^{\xi}$ and $\alpha^T$ with $0<{\xi}<T$.
We omit the proof of the following fact which can be verified by a
simple calculation: relation (\ref{Eq C^s via C^T}) implies that kernels $k^{\xi}$
and $k^T$ are connected as
\begin{equation*}
k^{\xi}(t,s)=k^T(t+(T-{\xi}),s+(T-{\xi})), \qquad 0\leqslant
t,s\leqslant {\xi}.
\end{equation*}
The latter leads to
\begin{equation}\label{Eq k^s via k^T}
k^{\xi}({\xi}-t,{\xi}-s)=k^T(T-t,T-s), \qquad 0\leqslant t,s\leqslant
{\xi}.
\end{equation}
Defining
$$
\hat k^T(t,s):=k^T(T-t,T-s),\qquad 0\leqslant t,s\leqslant T,
$$
relation (\ref{Eq k^s via k^T}) takes the form
$$
\hat k^{\xi}(t,s)=\hat k^T(t,s), \qquad 0\leqslant
t,s\leqslant \xi.
$$
Thus the function $\hat k^T$ does not depend on the superscript $T$.
Redenoting $\hat k^T=: \hat k$, we conclude that the integral part of
the connecting operator necessarily has the form
\begin{equation}\label{Eq form K^T}
(K^T\phi)(t)=\int_0^T\hat k(T-t,T-s)\,\phi(s)\,ds, \qquad
0\leqslant t\leqslant T,
\end{equation}
with a function $\hat k\in C^\infty([0,\infty)\times[0,\infty))$. Systems
$\alpha$ and $\alpha^T$, as well as all their attributes
(spaces and operators), are determined by the operator
$L_0$. Therefore, the following can be regarded as an assumption on this operator which
supplements Conditions \ref{C1} and \ref{C2}.

\begin{Condition}\label{C3}
For all $T>0$ the (scalarized) connecting operator $\dot C^T$ of
the system $\alpha^T$ takes the form (\ref{Eq C=I+K}), (\ref{Eq
form K^T}) and is an isomorphism in $\dFT$.
\end{Condition}

 In this formulation it is meant that the kernels of all
$K^T$ are determined by the same function $\hat k$. Also, since $\dot C^T-\mathbb I$ is compact in ${L_2(0,T)}$, Condition \ref{C3} implies that $\dot C^T$ is an isomorphism, which, owing to the equality $\dCT=(\dWT)^*\dWT$, in turn implies that operators $\dot W^T$ and $W^T$ are isomorphisms for each $T>0$. 

Note that some conditions are known which enable one to realize an
operator in a Hilbert space as an integral operator
\cite{Kor}. Therefore, in principle, Condition \ref{C3} is
efficiently checkable.

\subsubsection*{Main result}

\bul Denote $\tH:={L_2(0,\infty)}$; let $q$ be a smooth real-valued function (potential) on $[0,\infty)$. The minimal Schr\"{o}dinger operator $S_0=S_0^q$ associated with $q$ is
the closure of the operator $(-D^2+q)\upharpoonright C^\infty_{\rm c}(0,\infty)$. We define a class ${\cal Q}$ of smooth potentials by the conditions that extension of $q$ by zero to $\mathbb R$ preserves smoothness and that the operator $S_0$ is positive definite. Equivalently, preservation of smoothness means that all derivatives of $q$ at zero vanish, $q^{(j)}(0)=0$ for  $j\geqslant 0$. Note that positive definiteness of the minimal Schr\"odinger operator $S_0$ by the Glazman--Povzner--Wienholtz theorem \cite{DM, Hart} implies that 
$q$ is in the limit point case at infinity. Then $S_0$ has defect indices $n_{\pm}(L_0)=1$.
\smallskip

Our main result is the following.
\begin{Theorem}\label{T1}
Let $L_0$ be a closed symmetric positive definite operator in a Hilbert	space. If $L_0$ satisfies Conditions \ref{C1}, \ref{C2} and \ref{C3}, then there exists a potential $q\in{\cal Q}$ such that $L_0$ is unitarily equivalent to the minimal Schr\"odinger operator $S_0=S_0^q$.
\end{Theorem}

\section{Proof of Theorem \ref{T1}}

\subsubsection*{Triangular factorization}
Define in $\dot{\mF}^T$ the operator of inversion $Y^T:f(t)\mapsto f(T-t)$ (which is unitary, self-adjoint and satisfies $(Y^T)^2=\mathbb I$) and the operator
$\hat C^T:=Y^T\dot C^TY^T$,
$$
(\hat C^Tf)(t)=f(t)+\int_0^T\hat k(t,s)f(s)ds.
$$

We use the following factorization result by M. G. Krein \cite{GK}. Since the kernel $\hat k$ is continuous in $[0,T]^2$ and the operator $\hat C^{\xi}$ has a bounded inverse defined on the whole space $\dot{\mF}^{\xi}$ for every $\xi\in(0,T]$, the operator $(\hat C^T)^{-1}$ has a \emph{triangular factorization}
$$
(\hat C^T)^{-1}=(\mathbb I+V_+^T)(\mathbb I+V_-^T),
$$
where the kernels have the following properties:
$$
\begin{array}{l}
V_+^T(t,s)=0\text{ for }a\leqslant s<t\leqslant b,\quad \text{(a \emph{left Volterra operator)}},
\\
V_-^T(t,s)=0\text{ for }a\leqslant t\leqslant s\leqslant b, \quad \text{(a \emph{right Volterra operator)}}.
\end{array}
$$
Moreover, if one writes $(\hat C^{\xi})^{-1}=\mathbb I+\Gamma^{\xi}$ where
$$
(\Gamma^{\xi}f)(t)=\int_0^{\xi}\Gamma^{\xi}(t,s)f(s)ds
$$
with $\Gamma^{\xi}\in C([0,\xi]^2)$, then
$$
V_+^T(t,s)=V_+(t,s)=\left\{
\begin{array}{cl}
\Gamma^s(t,s),\quad &0\leqslant t\leqslant s\leqslant T,
\\
0,\quad &0\leqslant s<t\leqslant T,
\end{array}
\right.
$$
$$
V_-^T(t,s)=V_-(t,s)=\left\{
\begin{array}{cl}
0,\quad &0\leqslant t<s\leqslant T,
\\
\Gamma^t(t,s),\quad &0\leqslant s\leqslant t\leqslant T,
\end{array}
\right.
$$
Using standard arguments of the integral equations theory one can show that $\Gamma^{\xi}(t,s)$ is smooth in the square $[0,\xi]^2$ (is of the same smoothness as $\hat k$). We see that the kernels $V_{\pm}^T(t,s)$ do not depend on $T$ and are smooth in the corresponding triangles. Since $(\hat C^T)^{-1}\geqslant0$,
$$
(\hat C^T)^{-1}=((\hat C^T)^{-1})^*=(\mathbb I+(V_-^T)^*)(\mathbb I+(V_+^T)^*),
$$
and owing to uniqueness of factorization (see \cite{GK}) we have $(V_+^T)^*=V_-^T$. Denote
$$
B^T:=V_+^T,\quad A^T:=(\mathbb I+V_+^T)^{-1}-\mathbb I=-B^T+(B^T)^2-(B^T)^3+\cdots,
$$
and $A^T$ is also a left Volterra operator with a smooth kernel $a(t,s)$ which does not depend on $T$, that is
$$
\begin{array}{l}
((\mathbb I+A^T)f)(t)=f(t)+\int_t^Ta(t,s)f(s)ds,
\\
((\mathbb I+B^T)f)(t)=f(t)+\int_t^Tb(t,s)f(s)ds.
\end{array}
$$
Then
$(\hat C^T)^{-1}=(\mathbb I+B^T)(\mathbb I+B^T)^*$, $\hat C^T=(\mathbb I+A^T)^*(\mathbb I+A^T)$,
$$
\dot C^T=Y^T\hat C^TY^T=((\mathbb I+A^T)Y^T)^*(\mathbb I+A^T)Y^T=(\til W^T)^*\til W^T,
$$
where
$$
\til W^T:=(\mathbb I+A^T)Y^T.
$$
Summing up, we have proved the following result.

\begin{Lemma}
	For every $T>0$ operator $\dot C^T$ has a factorization $\dot C^T=(\til W^T)^*\til W^T$, where
	$$
	(\til W^Tf)(t)=f(T-t)+\int_t^Ta(t,s)f(T-s)ds,\quad t\in[0,T],
	$$
	with $a\in C^{\infty}(\{(t,s)\in[0,\infty)^2\ver s\leqslant t\})$.
\end{Lemma}

Consider $\til{\mathscr H}^T:=L_2(0,T)$ as a subspace of $\til {\mathscr H}$. Operator $\til W^T$ acts from $\dot \mF^T$ to $\til{\mathscr H}^T$, so that $\Ran\til W^T=\til{\mathscr H}^T$ and $(\til W^T)^{-1}$ is a bounded operator from $\til{\mathscr H}^T$ to $\dot \mF^T$.

\begin{Lemma}\label{lem unitary on H^T}
	$\Phi^T:=\til W^T(\dot W^T)^{-1}$ is a unitary operator from $\overline{\mU^T}$ to $\til{\mathscr H}^T$.
\end{Lemma}

\begin{proof}
	We have $\dot C^T=(\dot W^T)^*\dot W^T=(\til W^T)^*\til W^T$. This does not contradict uniqueness of factorization, because $\dot W^T$ and $\til W^T$ map to different spaces, but this means that they differ by a unitary operator:
	\begin{multline*}
	\Phi^T=\til W^T(\dot W^T)^{-1}=((\til W^T)^*)^{-1}\dot C^T(\dot W^T)^{-1}=((\til W^T)^*)^{-1}(\dot W^T)^*
	\\=(\dot W^T(\til W^T)^{-1})^*
	=((\til W^T(\dot W^T)^{-1})^{-1})^*=((\Phi^T)^{-1})^*.
	\end{multline*}
\end{proof}

Consider the sets
$$
\til\mU^T:=\Phi^T\mU^T=\til W^T\mM^T
$$
and
$$
\til\mU:={\rm span\,}\{\til u\in\til\mU^T\ver T>0\}.
$$
On the one hand, since for $\til u\in\til\mU^T$ there exists $f\in\mM^T=C_{\rm c}^{\infty}(0,T]$ such that $\til u(t)= (\til W^Tf)(t)=f(T-t)+\int_t^Ta(t,s)f(T-s)ds$, it follows that $\til\mU^T\subset C_{\rm c}^{\infty}[0,T)$. On the other hand, for $\til u\in C_{\rm c}^{\infty}[0,T)$ one has $((\til W^T)^{-1}\til u)(t)=(Y^T(I+B^T)\til u)(t)=\til u(T-t)+\int_{T-t}^Tb(T-t,s)\til u(s)ds\in C_{\rm c}^{\infty}(0,T]=\mM^T$, and it follows that $C_{\rm c}^{\infty}[0,T)\subset \til W^T\mM^T=\til\mU^T$. Thus 
$$
\til\mU^T=C^{\infty}_{\rm c}[0,T), \quad T>0,
$$
and 
$$
\til\mU=C_{\rm c}^{\infty}[0,\infty).
$$

\begin{Lemma}
	There exists a unitary operator $\Phi:{\mathscr H}\to\til {\mathscr H}$ such that for every $T>0$ one has $\Phi\upharpoonright{\overline{\mU^T}}=\Phi^T$.
\end{Lemma}

\begin{proof}
	According to \eqref{steady-state}, $\dot W^T\STX=\dot W^{\xi}$, $0<\xi<T$. Let us see that also $\til W^T\STX=\til W^{\xi}$: $(\til W^T\STX f)(t)=((\mathbb I+A^T)\til f)(t)$, where
	$$
	\til f(t)=\left\{
	\begin{array}{cl}
	f(\xi-t),&t\in[0,\xi],
	\\
	0,&t\in(\xi,T],
	\end{array}
	\right.
	$$
	so that indeed
	$$
	(\til W^T\STX f)(t)=f(\xi-t)+\int_t^{\xi}a(t,s)f(\xi-s)ds=(\til W^{\xi}f)(t).
	$$
	Consider $\Phi^T\upharpoonright{\overline{\mU^{\xi}}}=\til W^T(\dot W^T)^{-1}\upharpoonright{\overline{\mU^{\xi}}}$. Since $\dot W^T\STX=\dot W^{\xi}$, we have 
	$$
	\dot W^T\STX(\dot W^{\xi})^{-1}=\mathbb I_{{\overline{\mU^{\xi}}}}.
	$$
	Then
	$$
	\Phi^T\upharpoonright{\overline{\mU^{\xi}}}=\til W^T(\dot W^T)^{-1}\dot W^T\STX(\dot W^{\xi})^{-1}=\til W^T\STX(\dot W^{\xi})^{-1}=\til W^{\xi}(\dot W^{\xi})^{-1}=\Phi^{\xi}.
	$$
	For every $v\in\mU$ (a dense set in ${\mathscr H}$) there exists $T>0$ such that $v\in \mU^T$, so we define $\Phi_0 v:=\Phi^Tv$, and owing to the above relation this definition does not depend on the choice of $T$. By Lemma \ref{lem unitary on H^T} the closure $\Phi$ of the operator $\Phi_0$ is defined on the whole space ${\mathscr H}$ and is isometric. But since
	\begin{equation*}
	\Ran\Phi_0=
	{\rm span\,}\{\Phi_0(\mU^T)\ver T>0\}={\rm span\,}\{\til\mU^T\ver T>0\}=\til\mU=C_{\rm c}^{\infty}[0,\infty)
	\end{equation*}
	is dense in ${L_2(0,\infty)}=\til {\mathscr H}$, the operator $\Phi$ is unitary. From the fact that $\Phi^T\upharpoonright{\mU^T}=\Phi_0\upharpoonright\mU^T=\Phi^T\upharpoonright\mU^T$ it follows that $\Phi\upharpoonright{\overline{\mU^T}}=\Phi^T\upharpoonright{\overline{\mU^T}}=\Phi^T$ for every $T>0$.
\end{proof}

\subsubsection*{Operator $S^+_0$}

Let $T>0$, $f\in \mM^T$. From \eqref{Eq Wd=dW} we see that the operator $L_0^*\upharpoonright{\mU^T}$ in $\overline{\mU^T}$ is similar to the operator $-d^2/dt^2\upharpoonright{\mM^T}$ in $\mF^T$:
$$
\LUT=\dot W^T(-D^2\upharpoonright\mM^T)(\dot W^T)^{-1}
$$
Let us define the operator $\ST$ in $\til{\mathscr H}^T$,
$$
\Dom\ST=\til{\mU}^T,
\quad
\ST:=\til W^T(-D^2\upharpoonright\mM^T)(\til W^T)^{-1}.
$$
We have
\begin{multline*}
\ST=\til W^T(\dot W^T)^{-1}(\LUT)\dot W^T(\til W^T)^{-1}=\Phi^T(\LUT)(\Phi^T)^{-1}\\=\Phi(\LUT)\Phi^{-1}.
\end{multline*}
Since $\mU={\rm span\,}\{u\in\mU^T\ver T>0\}$, define the operator $\SO$ in $\til{\mathscr H}$,
$$
\Dom\SO=\til{\mU},
\quad
\SO:=\Phi(L_0^*\upharpoonright{\mU})\Phi^{-1},
$$
so that $\ST=\SO\upharpoonright{\til{\mU}^T}$.

\begin{Lemma}
	Operator $\SO$ acts as
	$$
	(\SO\til u)(t)=-\til u''(t)+q(t)\til u(t)+\int_t^{+\infty}Q(t,s)\til u(s)ds,\quad \til u\in\til{\mU},
	$$
	where 
	\begin{equation}\label{q}
	q(t)=2\left(\frac d{dt}(b(t,t))+b^2(t,t)\right)
	\end{equation}
	and $Q\in C^{\infty}(\{(t,s)\in[0,\infty)^2\ver t\leqslant s\})$.
\end{Lemma}

\begin{Remark}
	For every $\til u\in\til{\mU}=C_{\rm c}^{\infty}[0,\infty)$ there exists $T>0$ such that $\til u\in \til{\mU^T}=C_{\rm c}^{\infty}[0,T)$ and $\int_t^{+\infty}Q(t,s)\til u(s)ds=\int_t^TQ(t,s)\til u(s)ds$.
\end{Remark}

\begin{proof}
	Let $\til u\in\til{\mU}^T$. Then
	\begin{multline*}
	\SO\til u=\til W^T(-D^2\upharpoonright\mM^T)(\til W^T)^{-1}\til u
	=(\mathbb I+A^T)Y^T(-D^2\upharpoonright\mM^T)Y^T(\mathbb I+B^T)\til u
	\\=(\mathbb I+A^T)(-D^2\upharpoonright\mM^T)(\mathbb I+B^T)\til u,
	\end{multline*}
	and after a tedious, but straightforward calculation involving integration by parts and changing the order of integration, using $\til u(T)=\til u'(T)=0$ one gets:
	\begin{multline*}
	(\SO\til u)(t)=-\til u''(t)-\int_t^T(D_1^2b)(t,s)\til u(s)ds
	\\
	+(2(D_1b)(t,t)+(D_2b)(t,t))\til u(t)+b(t,t)\til u'(t)
	\\
	-\int_t^T ds\, a(t,s)\int_s^T(D_1^2b)(s,p)\til u(p)dp
	\\
	+\int_t^Ta(t,s)(-\til u''(s)+2(D_1b)(s,s)\til u(s)+(D_2b)(s,s)\til u(s)+b(s,s)\til u'(s))ds
	\\=-\til u''(t)+(a(t,t)+b(t,t))\til u'(t)+q(t)\til u(t)+\int_t^TQ(t,s)\til u(s)ds,
	\end{multline*}
	where
	$$
	q(t):=2(D_1b)(t,t)+(D_2b)(t,t)-(D_2a)(t,t)-a(t,t)b(t,t),
	$$
	\begin{multline*}
	Q(t,s):=-(D_2^2a)(t,s)-(D^2_1b)(t,s)-(D_2a)(t,s)b(s,s)
	\\+a(t,s)(D_1b)(s,s)-\int_t^sa(t,\tau)(D_1^2b)(p,s)dp.
	\end{multline*}
	Since $(\mathbb I+A)(\mathbb I+B)=\mathbb I$, or $A+B+AB=0$, the kernels $a$ and $b$ should satisfy the identity
	$$
	a(t,s)+b(t,s)+\int_s^ta(t,p)b(p,s)dp\equiv0.
	$$
	Putting $s=t$ we get that $a(t,t)+b(t,t)\equiv0$. Using this we get \eqref{q}, which completes the proof.
\end{proof}

\noindent Since one has
\begin{multline*}
{\rm Ker\,}\Gamma_1={\rm Ker\,}(L^{-1}L_0^*-\mathbb I)=\{u\in\Dom L_0^*\ver u=L^{-1}L_0^*u\}
\\=\{u\in\Dom L\ver Lu=L_0^*u\}=\Dom L,
\end{multline*}
it follows that
$$
\mU^T\cap\Dom L=\{u^f(T)\ver f\in\mM^T,\Gamma_1u^f(T)=0\}
=\{u^f(T)\ver f\in\mM^T,f(T)=0\}.
$$
Denote
$$
\mM_0^T:=\{f\in\mM^T\ver f(T)=0\},
$$
and
$$
\mU_0^T:=\mU^T\cap\Dom L=W^T\mM_0^T,
$$
$$
\mU_0:=\mU\cap\Dom L={\rm span\,}\{u\in\mU_0^T\ver T>0\}.
$$
Consider as well
$$
\til{\mU}_0^T:=\Phi\mU_0^T=\til W^T\mM_0^T,
$$
$$
\til{\mU}_0:=\Phi\mU_0.
$$
As before, owing to the relations 
$$
\til{\mU}^T=(\mathbb I+A^T)Y^T\mM^T,\quad
\mM^T=Y^T(\mathbb I+B^T)\til{\mU}^T,
$$
$$
\til u(t)=f(T-t)+\int_t^Ta(t,s)f(T-s)ds,
\quad
f(t)=\til u(T-t)+\int_{T-t}^Tb(T-t,s)\til u(s)ds,
$$
one can write
$$
\til{\mU}^T_0=\{\til u\in C_{\rm c}^{\infty}[0,T)\ver \til u(0)+\int_0^Tb(0,s)\til u(s)ds=0 \}
$$
and 
$$
\til{\mU}_0=\{\til u\in C_{\rm c}^{\infty}[0,\infty)\ver \til u(0)+\int_0^{\infty}b(0,s)\til u(s)ds=0 \}.
$$
Since $\mM_0^T$ is dense in $\mF^T$, the same is true for $\til{\mU}^T_0$ in $\til{\mathscr H}^T$ and for $\mU^T_0$ in $\overline{\mU^T}$.

\begin{Lemma}
One has $b(0,\cdot)\equiv0$, $\Im q=0$ and $Q=0$.
\end{Lemma}

\begin{proof}
	Let $T>0$. The operator $L_0^*\upharpoonright{\mU_0^T}=L_0^*\upharpoonright{(\mU^T\cap\Dom L)}=L\upharpoonright{(\mU^T\cap\Dom L)}$ is symmetric, and hence $\SO\upharpoonright{\til{\mU}_0^T}=\Phi (L_0^*\upharpoonright{\mU_0^T})\Phi^{-1}$ is also symmetric. Denote $S_1:=\SO\upharpoonright{\til{\mU}_0^T}$, $a:=\overline{b(0,\cdot)}\in C^{\infty}[0,T]$. Then
	$$
	\Dom S_1=\{u\in C_{\rm c}^{\infty}[0,T)\ver u(0)+(u,a)=0\}.
	$$
	Let us prove that the adjoint of $S_1$ in $L_2(0,T)$ is the operator
	$$
	S_2u:=-u''+\overline q u+Q^*u+u'(0)a,
	$$
	$$
	\Dom S_2:=\{u\in H^2(0,T)\ver u(0)=0\},
	$$
	where
	$$
	(Q^*u)(t):=\int_0^t\overline{Q(s,t)}u(s)ds.
	$$
	We do this in three steps.
	\\
	1. First observe that $\overline S_1u=-u''+qu+Qu$,
	$$
	\Dom\overline S_1=\{u\in H^2(0,T)\ver u(0)+(u,a)=u(T)=u'(T)=0\}.
	$$
	\\
	2. Let us prove that $\overline S_1\subset S_2^*$.
	\\
	Pick $u\in H^2(0,T)$ such that $u(0)+(u,a)=u(T)=u'(T)=0$. For every $\varphi\in\Dom S_2$ one has (since $u(T)=u'(T)=0$ and $u(0)+(u,a)=0,\varphi(0)=0$)
	\begin{multline*}
	(S_2\varphi,u)=\int_0^T(-\varphi''+\overline q \varphi+Q^*\varphi)\overline u+\varphi'(0)(a,u)
	\\
	=\int_0^T\varphi(\overline{-u''+qu+Qu})+\varphi'(0)(\overline{u(0)+(u,a)})-\varphi(0)\overline{u'(0)}
	\\
	=\int_0^T\varphi(\overline{-u''+qu+Qu}),
	\end{multline*}
	hence $u\in\Dom S_2^*$ and $S_2^*u=-u''+qu+Qu$.
	\\
	3. Let us prove that $S_2^*\subset\overline S_1$.
	\\
	Pick $u\in\Dom S_2^*$. For every $\varphi\in\Dom S_2$ one has $(S_2\varphi,u)=(\varphi,h)$, where $h=S_2^*u$. In particular, for every $\varphi\in C_{\rm c}^{\infty}(0,T)$
	\begin{multline*}
	(S_2\varphi,u)=\int_0^T(-\varphi''+\overline q\varphi+Q^*\varphi+\varphi'(0)a)\overline u
	\\
	=\int_0^T-\varphi''\overline u+\int_0^T\varphi(\overline{qu+Qu})=\int_0^T\varphi\overline h,
	\end{multline*}
	or
	$$
	\int_0^T\varphi''\overline u=\int_0^T\varphi(\overline{qu+Qu-h}),\quad\forall\varphi\in C_{\rm c}^{\infty}(0,T).
	$$
	It follows that there exists the generalized second derivative $u''\in L_2(0,T)$, or that $u\in H^2(0,T)$.
	
	For every $\varphi\in\Dom S_2$
	\begin{multline*}
	(S_2\varphi,u)=\int_0^T(-\varphi''+\overline q\varphi+Q^*\varphi+\varphi'(0)a)\overline u
	\\
	=\int_0^T\varphi(\overline{-u''+qu+Qu})+\varphi'(0)(a,u)-(\varphi'\overline u-\varphi\overline{u'})\upharpoonright0^T
	\\
	=\int_0^T\varphi(\overline{-u''+qu+Qu})+\varphi'(0)(\overline{u(0)+(u,a)})
	\\+\varphi(T)\overline{u'(T)}-\varphi'(T)\overline{u(T)}
	=\int_0^T\varphi\overline h.
	\end{multline*}
	Hence for every $\varphi\in\Dom S_2$ one has
	$$
	\int_0^T\varphi(\overline{-u''+qu+Qu-h})=\varphi'(0)(\overline{u(0)+(u,a)})+\varphi(T)\overline{u'(T)}-\varphi'(T)\overline{u(T)}.
	$$
	One can choose sequences $\{\varphi_n\}_{n\in\mathbb N}\subset\Dom S_2$ such that $\varphi_n\to0$ in $L_2(0,T)$ and the values $\varphi_n'(0)$, $\varphi_n(T)$ and $\varphi'_n(T)$ are constant. Since these constants can be arbitrary and the value of the left-hand side goes to zero as $n\to\infty$, it has to be
	$$
	u(0)+(u,a)=u(T)=u'(T)=0.
	$$
	Therefore $u\in\Dom\overline S_1$.
	
	From 2. and 3. it follows that $S_1^*=S_2$. Since $S_1$ is symmetric, one must have $S_1\subset S_2$. This means that, firstly, $\Dom S_1\subset\Dom S_2$ and, secondly, $S_1u=S_2u$ for every $u\in\Dom S_1$. From the first fact it follows that $u(0)=0$ for every $u\in\Dom S_1$ and hence $u\perp a$, so that $\Dom S_1\subset a^{\perp}$. However, $\Dom S_1$ is dense in $L_2(0,T)$, and $\Dom S_1\subset a^{\perp}$ can hold, only if $a=0$. From the second fact it follows that $qu+Qu=\overline q u+Q^* u$ for every $u\in\Dom S_1$. This implies equality of the self-adjoint operator of multiplication by the smooth function  $\Im q$ and the compact operator $\Im Q=(1/2i)(Q-Q^*)$. This can only be true, if the spectrum of both of them consists of a single point zero, which means that $\Im q\equiv0$ and $\Im Q=0$. Since $Q$ is a right integral Volterra operator and $Q^*$ is a left one, from $Q-Q^*=0$ it follows that $Q=Q^*=0$. The proof of the lemma is complete.
\end{proof}

\subsubsection*{Completing the proof of Theorem \ref{T1}}
Since $b(0,\cdot)\equiv0$, one can write
$$
\til{\mU}_0=\{\til u\in C_{\rm c}^{\infty}[0,\infty)\ver \til u(0)=0 \}.
$$
We see that $\SO\upharpoonright{\til{\mU}_0}$ acts as a Schr\"odinger operator on its domain, it is symmetric and positive definite, its domain contains $C_{\rm c}^{\infty}(0,\infty)$. Therefore the minimal Schr\"odinger operator $S_0=S_0^q\subset \SO\upharpoonright{\til{\mU}_0}$ (defined as the closure of  $-D^2+q$ on $C_{\rm c}^{\infty}(0,\infty)$) is also positive definite. By the Glazman--Povzner--Wienholtz theorem \cite{DM, Hart} it follows that the potential $q$ is in the limit point case at infinity. Then $S_0$ has defect indices $n_+(S_0)=n_-(S_0)=1$. One has
	\begin{multline*}
	\Dom \overline{\SO}=
	\{\til u\in{L_2(0,\infty)}\ver\til u,\til u'\text{ are locally a. c., }
	\\
	-\til u''+q\til u\in{L_2(0,\infty)}\}=\Dom S_0^*,
	\end{multline*}
	which means that $\overline{\SO}= S_0^*$. We have the following chain of inclusions:
	$$
	\Phi L_0\Phi^*\subset(\SO)^*=S_0\subset S_0^*=\overline{\SO}\subset\Phi L_0^*\Phi^*.
	$$
	Since $n_{\pm}(L_0)=1$ and $S_0\neq S_0^*$, it should be that
	$$
	\Phi L_0\Phi^*=(\SO)^*=S_0\quad\text{and}\quad S_0^*=\overline{\SO}=\Phi L_0^*\Phi^*.
	$$
	So we arrive at the equality $\Phi L_0\Phi^*=S_0$.

	It remains to show that $q^{(j)}(0)=0$ for all $j\geqslant0$. Note that connecting operators $\dot C^T$ for unitarily equivalent $L_0$ and $S_0$ coincide, if we choose the element $\Phi e$ for the scalarization of $C^T$ for $S_0$. So we consider the system $\alpha^T$, which corresponds to $S_0$ in ${\mathscr H}={L_2(0,\infty)}$. It is known that its connecting operator has the form
	$$
	(\dot C^T\phi)(t)=\phi(t)+\int_0^T\left[\,p(2T-t-s)-p(|t-s|)\,\right]\phi(s)\,ds,
	$$
	where $p(t)={1\over 2}\,\int_0^tr(s)\,ds$ and $r$ is the
	so-called response function, which plays the role of inverse data in
	the classical time-domain Sturm--Liouville inverse problem (see,
	e.g., \cite{BMikh}). For smooth potentials $q$ the function $r$ is also smooth, but, owing to the presence of $|t-s|$, the kernel of the integral operator $\dot C^T-\mathbb I$ may be nonsmooth at the diagonal $t=s$. This is not the case, if and only if $r(t)$ vanishes at $t=0$
	along with  all of its derivatives. One can show that $r^{(j)}(0)=0$, $j\geqslant0$, is equivalent to the condition that $q^{(j)}(0)=0$, $j\geqslant0$. This completes the proof of Theorem \ref{T1}.

\subsubsection*{Comments}
\bul In fact, something more is proved by Theorem \ref{T1}: it
provides a characterization of a class of operators. Namely, given
a closed symmetric positive definite operator $L_0$ in a Hilbert space, it is unitarily equivalent to the minimal
Schr\"odinger operator $S_0^q$ with some potential $q\in {\cal Q}$, if and {\it
only if}\,\, Conditions \ref{C1}, \ref{C2} and \ref{C3} are satisfied.
Indeed, owing to unitary equivalence of $L_0$ and $S_0^q$ and to coincidence of their $\dot C^T$, it suffices to show that $S_0^q$ itself satisfies these three conditions, if $q\in{\cal Q}$. Conditions \ref{C1} and \ref{C2} are satisfied for any $q$ in the limit point case at infinity. The argument in the last paragraph of the proof of Theorem \ref{T1} can be used to see that Condition \ref{C3} is also satisfied.

\bul It would be reasonable to expect similar results in the case
$0<n_\pm(L_0)<\infty$. We plan to deal with that in the future.

\section{Wave model}

\subsubsection*{Eikonal operator}

Let $P^t$ be the projection in ${\mathscr H}$ onto the reachable subspace $\overline{\mU^t}$. If $\overline{\mU}={\mathscr H}$, then the function $P^t$ (continued by zero to the values $t\leqslant0$) is a resolution of identity and defines the spectral measure $dP^t$. With the system $\alpha$ of the form (\ref{alpha1})--(\ref{alpha3}) one associates the self-adjoint operator
$$
E:=\int_{[0,\infty)}t\,dP^t
$$
in ${\mathscr H}$. We call it the {\it eikonal} operator, or, shortly, the {\em eikonal}.

Under Condition \ref{C1} the eikonal operator can provide a functional model
of $L_0$ in the following way, which just realizes the spectral
theorem for self-adjoint operators (see \cite{BirSol}). Suppose the operator $E$ has simple spectrum and $e\in{\rm Ker\,} L_0^*$ is its generating vector. Note that the measure $(dP^t y,e)$ is absolutely continuous with respect to the measure $d\mu(t):=(dP^t e,e)$. To each $y\in {\mathscr H}$ one assigns a function
\begin{equation}\label{Eq eik realiz}
	y(t):=\frac{d(P^t y,e)}{d(P^t e,e)},\qquad t>0,
\end{equation}
which is an element of the space $\mE:=L_2([0,\infty),d\mu)$. The map $U_E:y\mapsto y(\cdot)$ is a unitary operator from ${\mathscr H}$ to $\mE$, it diagonalizes $E$: the operator $U_EEU_E^*$ is the operator of multiplication by the independent variable in $\mE$ (see \cite{AG}).

Respectively, operator $L_0$ is transformed into the operator
$$
L_0^{\rm w}:=U_EL_0U_E^*
$$
in $\mE$ which is referred to as the {\it wave model} of $L_0$
\cite{BSim_1}\footnote{In fact, the above construction is a	simplified version of the wave model, see \cite{BSim_1,BSim_3}.}. In other terms it can be said that the operator $L_0^{\rm w}$ is $L_0$ written in spectral representation of the eikonal operator.

\subsubsection*{System $\alpha$ for $S_0$}

\bul The system $\alpha$ associated with the operator $S_0=S_0^q$ in ${L_2(0,\infty)}$ with a
potential $q\in{\cal Q}$ is
\begin{align}
	\label{alpha1S} & u_{tt}(x,t)-u_{xx}(x,t)+q(x)u(x,t)=0,  && x>0,\ t>0,\\
	\label{alpha2S} & u(x,0)=u_t(x,0)=0, && x\geqslant 0,\\
	\label{alpha3S} & u(0,t) = f(t), && t \geqslant 0.
\end{align}
Its inner space is ${\mathscr H}=L_2(0,\infty)$, its reachable sets are \cite{BSim_1}
$$
\mU^T=C_{\rm c}^{\infty}[0,T),\quad \overline{\mU^T}=L_2(0,T),\quad T>0.
$$
Respectively, projections on $\overline{\mU^T}$ are operators $P^T$ which
multiply functions from ${L_2(0,T)}$ by interval indicators:
$P^Ty=\chi_{[0,T]}y$, i.e., cut off functions at the point $T$. As a
result, the eikonal operator of the system $\alpha$ acts as
$$
(Ey)(x)=\left(\int_{[0,\infty)}t\,dP^ty\right)(x)=x\,y(x),\qquad
x>0,
$$
i.e., is the operator of multiplication by independent variable in ${L_2(0,\infty)}$.
\smallskip

\bul Since $q\in{\cal Q}$ is smooth and ${\rm Ker\,} S_0^*$ consists of those solutions to the equation $-\eps''+q\eps=0$ which belong to ${L_2(0,\infty)}$, the elements $\eps\in{\rm Ker\,} S_0^*$ are smooth functions (in fact, ${\rm Ker\,} S_0^*$ is a one-dimensional subspace and all its elements are proportional to $\eps$). They may have only isolated zeros, which can accumulate only at infinity. As one can see, $\eps$ is a generating element of the operator $E$.

Realizing elements of ${\mathscr H}$ using (\ref{Eq eik realiz}), we obtain
the operator
\begin{equation}\label{Eq eik realiz S0}
	U_E: \,y\mapsto y(\cdot),\,\,\, y(t):=\frac{d(P^t
		y,\eps)}{d(P^t \eps,\eps)}=\frac{y(t)}{\eps(t)},\qquad
	t>0,
\end{equation}
which maps ${\mathscr H}$ onto $\mE=L_2([0,\infty),d\mu)$ (with
$d\mu(t)=|\eps(t)|^2dt$). Respectively, we get the Sturm--Liouville operator
\begin{equation}\label{Eq wm of S_0}
S_0^{\rm w}=U_ES_0U_E^*
=\frac1{\eps}(-D^2+q)\eps
\end{equation}
on $\Dom S_0^{\rm w}=U_E\Dom S_0$ as the wave model of $S_0$.

\subsubsection*{The model}

Now let $L_0$ satisfy Conditions \ref{C1}, \ref{C2}, \ref{C3}, and let
$S_0=\Phi L_0\Phi^*$ be its unitary copy given by Theorem \ref{T1}. Fix a nonzero $e\in\mK={\rm Ker\,} L_0^*$ and let $\eps:=\Phi e\in{\rm Ker\,} S_0^*$. Let $S_0^{\rm w}$ be the wave model of $S_0$ constructed using the element $\eps$. Then, in view
of invariant character of the procedure that transforms an operator into
its wave model, the wave models of $L_0$ and $S_0$ turn out to be
identical and the copy $L_0^{\rm w}=S_0^{\rm w}$ is provided by the unitary
operator $\Phi^{\rm w}=U_E\Phi$, so that $L_0^{\rm w}=\Phi^{\rm w}L_0(\Phi^{\rm w})^*$ holds.

As a result we conclude that under assumptions of Theorem \ref{T1} the wave model of the operator $L_0$ is a Sturm--Liouville operator of the form (\ref{Eq wm of S_0}).


\begin{thebibliography}{99}
	
	\bibitem{AG}
	N. I. Akhiezer, I. M. Glazman. 
	Theory of linear operators in Hilbert space. 
	{\em Dover}, 
	1993.
	
	\bibitem{B_1D_BCm}
	M. I. Belishev. 
	\newblock{Boundary control and inverse problems: one-dimensional version of BC-method.} 
	{\em Journal of Mathematical Sciences}, 
	155(3), 2008, 343--378. 
	DOI:10.1007/s10958-008-9220-2.
	
	\bibitem{BArXiv}
	M. I. Belishev.
	\newblock {A unitary invariant of a semi-bounded operator in reconstruction
		of manifolds.}
	\newblock{\em Journal of Operator Theory}, 
	69(2), 2013, 299--326.
	
	\bibitem{DSBC_3}
	M. I. Belishev.
	\newblock {Wave propagation in abstract dynamical system with boundary
		control.} 
	arXiv:2307.00605v1, 2023.
	
	\bibitem{BD}
	M. I. Belishev, M. N. Demchenko.
	\newblock{Dynamical system with boundary control associated with a
		symmetric semibounded operator.}
	\newblock{\em Journal of Mathematical Sciences}, 
	194(1), 2013, 8--20.
	
	\bibitem{BMikh}
	M. I. Belishev, V. S. Mikhailov.
	\newblock{Unified approach to classical equations of inverse problem theory.}
	\newblock{\em Journal of Inverse and Ill-Posed Problems}, 
	20(4), 2012, 461--488.
	
	\bibitem{BSim_1}
	M. I. Belishev, S. A. Simonov.
	\newblock {Wave model of the Sturm--Liouville operator on the
		half-line.}
	\newblock{\em St. Petersburg Math. J.}, 
	29(2), 2018, 227--248.
	
	\bibitem{BSim_2}
	M. I. Belishev, S. A. Simonov.
	\newblock {A wave model of metric spaces.}
	\newblock{\em Functional Analysis and Its Applications},
	53(2), 2019, 79--85.
	DOI:10.1134/S0016266319020011.
	
	\bibitem{BSim_3}
	M. I. Belishev, S. A. Simonov.
	\newblock {A wave model of metric space with measure}.
	\newblock{\em Sbornik: 	Mathematics}, 
	211(4), 2020, 521--538.
	
	\bibitem{BSim_4}
	M. I. Belishev, S. A. Simonov.
	\newblock {On an evolutionary dynamical system of the first order with boundary control.}
	\newblock {\em Zapiski Nauchnykh Seminarov POMI}, 
	483, 2019, 41--54 (in Russian).
	
	\bibitem{BSim_5}
	M. I. Belishev, S. A. Simonov.
	\newblock {A canonical model of the one-dimensional dynamical Dirac system with boundary control.}
	\newblock {\em Evolution Equations and Control Theory}, 
	11(1), 2022, 283--300. 
	DOI:10.3934/eect.2021003.
	
	\bibitem{Blag}
	A. S. Blagovestchenskii.
	\newblock{On a local method of solution of a nonstationary inverse problem for a inhomogeneous string.}
	\newblock{\em Proc. Steklov Inst. Math.}, 
	115, 1971, 30--41.
	
	\bibitem{Blag2}
	A. S. Blagovestchenskii.
	\newblock{Inverse Problems of Wave Processes.}
	\newblock{\em VSP, Netherlands}, 
	2001.
	
	\bibitem{Birkhoff}
	G. Birkhoff.
	\newblock{Lattice Theory.}
	\newblock{\em Providence, Rhode Island}, 
	1967.
	
	\bibitem{BirSol}
	M. S. Birman, M. Z. Solomyak.
	\newblock {Spectral Theory of Self-Adjoint Operators in Hilbert Space.}
	\newblock {\em D. Reidel Publishing Comp.}, 
	1987.
	
	\bibitem{DM}
	V. A. Derkach, M. M. Malamud. 
	Extension theory of symmetric operators and boundary value problems. 
	{\em Proceedings of Institute of Mathematics of NAS of Ukraine}, 
	104, 2017.
	
	\bibitem{GK}
	I. Ts. Gohberg, M. G. Krein. 
	Theory and applications of Volterra operators in Hilbert space, 
	{\em AMS}, 
	1970.
	
	\bibitem{Hart}
	P. Hartman. 
	Differential equations with non-oscillatory eigenfunctions, 
	{\em Duke Math. J.}, 
	15, 1948, 697--709.
	
	\bibitem{Kato}
	T. Kato.
	Perturbation theory for linear operators.
	{\em Springer-Verlag}, 1966.
	
	\bibitem{Kor}
	V. B. Korotkov.
	\newblock{Integral operators.}
	\newblock{\em Novosibirsk: Nauka, SO AN SSSR}, 
	1983 (in Russian).
	
	\bibitem{Vishik}
	M. I. Vishik.
	\newblock {On general boundary value problems for elliptic differential equations.} 
	{\em Amer. Math. Soc. Transl. Ser.}, 
	224, 1963, 107--172.
	
\end{thebibliography}
\end{document}